# Prototype development of the Integral-Field unit for VIRUS

Andreas Kelz[*a], Svend M. Bauer[a], Frank Grupp[c], Gary J. Hill[b], Emil Popow[a],
Povilas Palunas[b], Martin M. Roth[a], Phillip J. MacQueen[b], Ute Tripphahn[a]

[a] Astrophysikalisches Institut Potsdam, An der Sternwarte 16, 14482 Potsdam, Germany
[b] McDonald Observatory, Univ. of Texas at Austin, 1 University Station, C1402 Austin, TX
[c] Universitäts-Sternwarte München, Scheinerstrasse 1, 81679 München, Germany

## ABSTRACT

VIRUS is a planned integral-field instrument for the Hobby-Eberly Telescope (HET). In order to achieve a large field-of-view and high grasp at reasonable costs, the approach is to replicate integral-field units (IFU) and medium sized spectrographs many times. The Astrophysical Institute Potsdam (AIP) contributes to VIRUS with the development and testing of the IFU prototype. While the overall project is presented by Hill et al.[1], this paper describes the opto-mechanical design and the manufacture of the fiber-based IFU subsystem. The initial VIRUS development aims to produce a prototype and to measure its performance. Additionally, techniques will be investigated to allow industrial replication of the highly specific fiber-bundle layout. This will be necessary if this technique is to be applied to the next generation of even larger astronomical instrumentation.

## 1. INTRODUCTION

Due its design, the Hobby-Eberly Telescope (HET), with a 9.2 meter effective aperture, but a segmented primary mirror, a fixed declination axis combined with a prime focus tracker, is most competitively used as a spectroscopic survey telescope. The future instrumentation plan for HET foresees a wide field of view corrector[2] and a highly multiplexed spectrograph[3]. A main science driver is the Hobby-Eberly Telescope Dark Energy Experiment[4] (HETDEX). HETDEX aims to map the 3D-spatial distribution of Lyα emitting galaxies (LAEs) over a volume of 5 Giga parsec[3]. This dataset is used to constrain the understanding of the expansion history of the Universe and to limit the parameter space about the evolution of dark energy.

In contrast to current emission-line surveys involving marrow band filters or Fabry-Perot etalons, it is of advantage to use a set of Integral Field (IF) spectrographs for this study. The concept for the Visible Integral-field Replicable Unit Spectrograph (VIRUS[1,3,5]), foresees around 145 IF spectrographs, covering a 22 arcmin. field. The advantage of an IF spectrograph for this project is that the tracer galaxies can be identified and have their redshifts determined in one observation. VIRUS is optimized to survey 250 square degrees at wavelengths from 340 to 570 nm, corresponding to redshifts between 1.8 and 3.7, with the expectation to find around 1 million LAEs in about 110 nights. Instead of building a classical monolithic instrument, which due to its size and costs is prohibitive in this case, the approach is to involve massive replication of a simple spectrograph that is connected to a fiber-based IF unit. While the concept of replication is also followed by other instrumental developments such as for MUSE[6], the replication rate and therefore the cost saving factor for VIRUS will be larger, as more than hundred identical copies of basic sub-systems will be realized.

The VIRUS development aims to produce a simple, inexpensive integral field spectrograph that can be built in "series" by commercial suppliers. The spectrograph itself is being designed at the University of Texas at Austin. The integral field unit (IFU) is based on a bare fiber-bundle design, in tradition with instruments such as DensePak[7], INTEGRAL[8], SparsePak[9], and PPak[10]. Based upon its experience with the Potsdam Multi-Aperture Spectrophotometer (PMAS[11]), the

---

[*] send correspondence to: akelz@aip.de; fax +49 331 7499-429

Astrophysical Institute Potsdam (AIP) was trusted with the development of the prototype of the integral-field unit and issues concerning the data reduction.

Section 2 of this paper summarizes the requirements for the field-unit. The opto-mechanical design, the manufacture, and the assembly of the prototype VIRUS-IFU is presented in section 3, 4 and 5, respectively. Various tests were undertaken to measure the performance of the optical fibers, which are described in section 6. Section 7 summarizes both the development up to date and future work.

## 2. REQUIREMENTS

### 2.1 IFU head format

Each VIRUS IFU consists of 246 fibers. Each fiber has a diameter of 200 microns and an area of 1.0 square arcseconds on the sky (1.13 arcsec. diameter) for an input beam of f/3.65. The fibers are arranged in a hexagonal close pack (HCP) with a square format as shown in Fig. 1, left. There are 17 rows of 14 or 15 fibers. The fill factor of the active area of the fibers will be close to 1/3 in order to provide maximum area coverage in three dithered exposures. There are different ways to achieve the required spacing for this fill-factor. Either fibers with the according core- and buffer sizes need to be obtained, or fibers can be inserted into precision glass capillary tubes with close tolerance on the inner diameters to match the outer diameters of the fibers, or into a precision aperture plate. The IFU bundle has one layer of buffer fibers around its outer edge to minimize variations in throughput due to focal ratio degradation. The fibers need to be polished and bonded to a thin fused silica plate with anti-reflection coating on its outer face, using a gel or liquid. The transmission of the couplant must be >99% over 340-670 nm, and the goal is to achieve better than 98% efficiency for the injection of light at f/3.65. The resultant fiber bundle has a size of approximately 5mm x 5mm.

The housing within which the IFU is mounted should have slim dimensions so as to allow the IFUs to be located next to each other as closely as possible. The requirement is for separations of twice the active IFU area, as to achieve a packing fraction with multiple IFUs as close as ¼. Shown in Fig. 1, right is a packing arrangement with 1/9 fill factor, which is a likely setup used for HETDEX. The IFU heads need to be held in place securely with features to fix position in all 3 spatial dimensions, orientation and rotation. No adjustment of position is needed, as CNC machine tolerance will be sufficient. The tolerance on these quantities is 0.1 mm in all dimensions. The focal surface of the new HET corrector is spherical with radius of curvature of about 1000 mm. It is concave to the direction of light propagation. The central axis of each IFU bundle will be held normal to the focal surface.

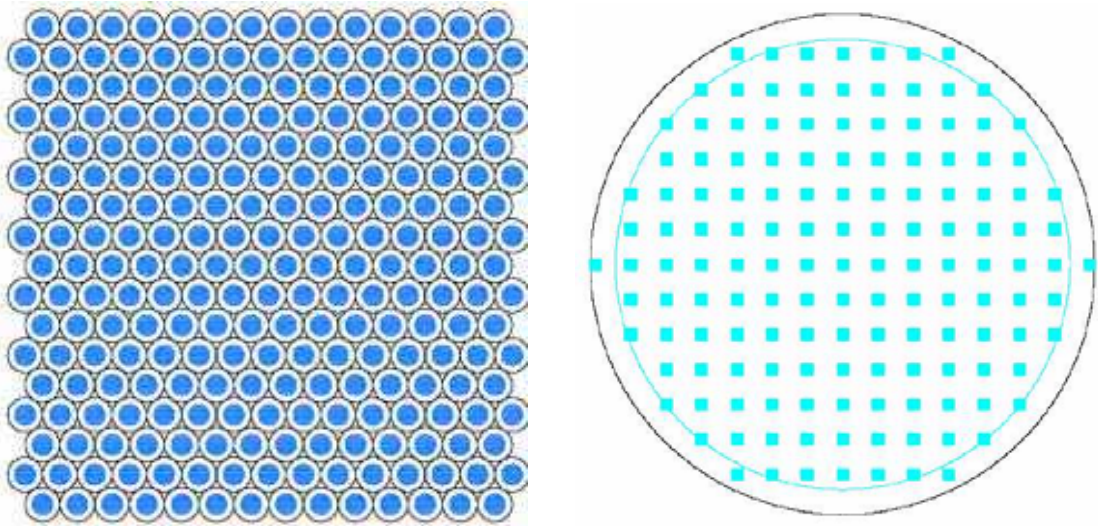

**Figure 1.** Left: Layout of active 200 µm cores of fibers (filled circles) in a single VIRUS IFU head. There are 17 rows of alternating 14 and 15 fibers. The fill factor is close to 1/3. The outer circles denote the capillary tubes used to set up the matrix with 340 µm pitch. Right: Possible layout of 143 IFUs (filled squares) for a full VIRUS. The fill factor is 1/9. The inner circle represents the 18 arcmin diameter science field and the outer circle the 20 arcmin diameter guide field of the new HET corrector.

## 2.2 Spectrograph input format of IFU

The rows of fibers at the IFU had are arrayed row-by-row at the input of the spectrograph in sets, with spacing between. The fibers are bonded to a monolithic substrate. For a stainless steel substrate, the change in position of the outer fibers is 17 microns for the required 35 degrees Celsius temperature range. The layout of the fibers is as follows: There are gaps of one missing fiber in 9 positions along the slit array to allow scattered light to be measured at the detector. There is a central gap with a missing fiber to allow two-amplifier read out of the detector, if this is needed. The input requirements are complex, as it is necessary that the fibers point normal to the surface of the collimating mirror at the appropriate field angle for their position in the input array. The fibers are immersed against the concave face of a cylindrical meniscus lens (radius of curvature close to 420 mm, concave) with optical couplant so as to minimize exit losses. The ends of the fiber banks must be polished curved, or the fibers must be pre-polished before assembly on the monolithic substrate.

The layout of 15 (out of 246) fibers is shown in Fig. 2. The fibers are aimed normal to a spherical surface with radius=420 mm so as to create an optical system with no preferred axis. The center-to-center separation of the fibers at the exit surface of the fibers is 320 microns with an angular pitch of 0.043 degrees. Misalignment of the fibers causes displacement of the pupil and is equivalent to FRD.

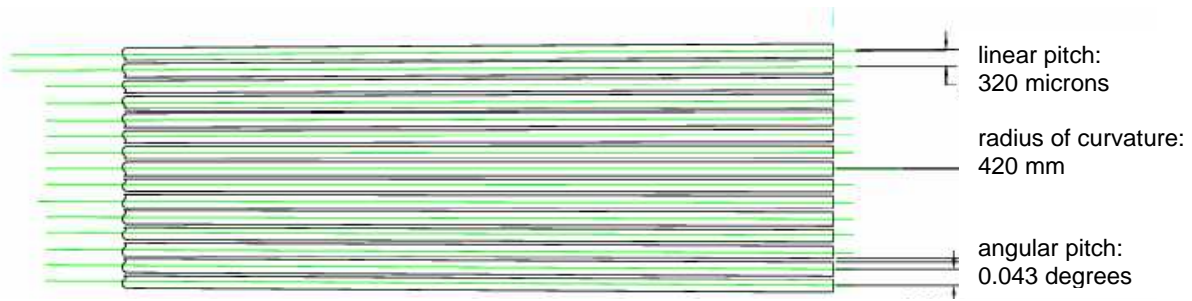

**Figure 2.** Layout of 15 fibers on the substrate. Note the change in angle needed to stay normal to the input surface with radius of 420 mm. The fibers are on 320 micron centers at this surface which translates to a nominal separation of about 70 microns at the detector. The fiber bank is polished, and then immersed against the 419.6 mm ROC input face of the cylindrical lens with optical couplant.

## 2.3 Fiber cable design

The length of the fiber run from HET focus to the VIRUS spectrograph modules was not fixed at the time of the prototype development. If the spectrographs can be placed close to the tracker, a minimum fiber length of 4 to 5 meters is possible. If the spectrographs are to be mounted at the side of the HET structure, the needed fiber length will increase to around 15 meters. The VIRUS module will be replicated >100 times and the bulk and weight of traditionally-sheathed IFU cables would be unwieldy. Hence, the majority of the fiber run will be made inside a single furcation tube. The exact design of the cable is yet to be determined, and will depend on the final location of the VIRUS spectrographs with respect to the IFU head. Each IFU and fiber cable must be independently removable from the telescope, cable wrap, and spectrograph.

## 2.4 Focal ratio degradation (FRD)

The input focal ratio to the bundle is f/3.65, so that a 200 micron fiber core corresponds to 1.13 arcseconds on the sky with the new wide field corrector for HET[2]. The collimator and camera of VIRUS can accept f/3.36, implying a maximum of 9% focal ratio degradation (for a description of FRD see Barden[12]) that is acceptable. This value includes the effect of FRD in the fiber, and also the pupil displacement due to fiber input angle errors (see above). The aim is to keep the effective FRD of all active fibers in the bundle below this level. The requirement is to have 95% of the light within a 125 mm diameter pupil, which corresponds to f/3.36, centered at the nominal optical axis. The optics produces a 115 mm diameter pupil in the absence of FRD.

## 3. OPTO-MECHANICAL DESIGN

All opto-mechanical design for the VIRUS-IFU was done at the Astrophysical Institute Potsdam (AIP) using the CAD software program Inventor®, version 10.

### 3.1 The IFU head mount

The IFU constructed at AIP is based on a matrix of fused silica capillary tubes arrayed in a hexagonal closed pack, where the outer diameters set the matrix spacing of the optical fibers that are inserted into the capillaries. A mount (Figure 3) was designed that is precise enough to ensure that the matrix is well aligned, but allows enough space to accommodate the diameter variations. Typically, the capillary diameter tolerance, as given by manufacturers, is 3 microns. Within one row of 15 capillaries, this could add up to 45 microns, which is enough for the matrix to get out of shape. On the other hand, a mount that is too tight causes problems to insert the last fibers and induces stress to the fibers, a potential source of FRD.

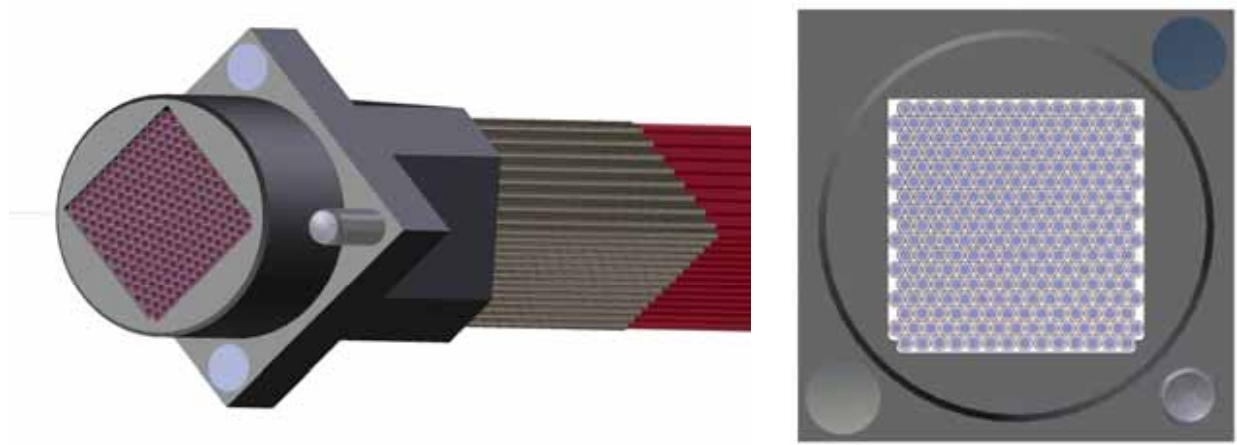

**Figure 3.** CAD view of a IFU head mount with a single fiber-bundle. The mount features magnetic buttons, a pin for re-producible positioning and measures only 9 mm on a side.

Additionally, the goal was to design a mount that can be easily, but precisely and securely plugged into a field-plate, so that various multi-IFU layouts are possible for the full VIRUS configuration (see Figure 4). While the design does not allow the IFUs to be buttable, a close packing fraction of ¼ can be achieved, as the overall mount measures less than twice the active fiber area.

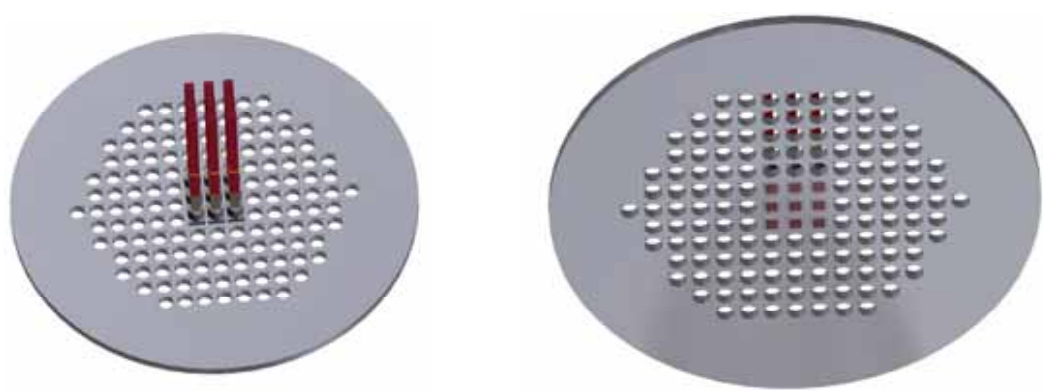

**Figure 4.** Top (left) and bottom (right) CAD view of a possible multi-IFU configuration, using a custom-made plug-plate. Nine (out of 143) IFU bundles are shown. The tightest packing fraction is as dense as ¼.

## 3.2 The slit-unit

The aim was to design a slit-unit (Fig. 5) that provides the necessary space to fan out the fibers from the narrow cable to the length of the slit, to provide some fiber-slack reservoir within the unit, while keeping the unit lightweight and compact. As the constraints for the fiber-slit placement are tight (see §2.2), the slit-unit mounts with precision pins to a reference plate and directly to the VIRUS collimator. The use of the reference plate ensures that any fiber-slit can be connected to any collimator in the full configuration, allowing easier maintenance and flexible usability in case of malfunction of subsystems.

The fiber slit itself is coupled to a cylindrical lens, which is the first optical element of the collimator. The individual fibers are cemented to a monolithic block that sets the spacing and angular pitch and is mounted to the lens cell. The cell aligns the optical axis of lens and fibers. A blackened aperture mask in front of the lens cell restricts the beam to be slower than f/1.19, and will prevent excess light, caused by FRD, to shine directly on the grating without being collimated. Also, collimated light that may reflect from the front of the slit-block and may appear as a ghost "slit" in the image is blocked by this slit mask and an additional mask at the collimator mirror.

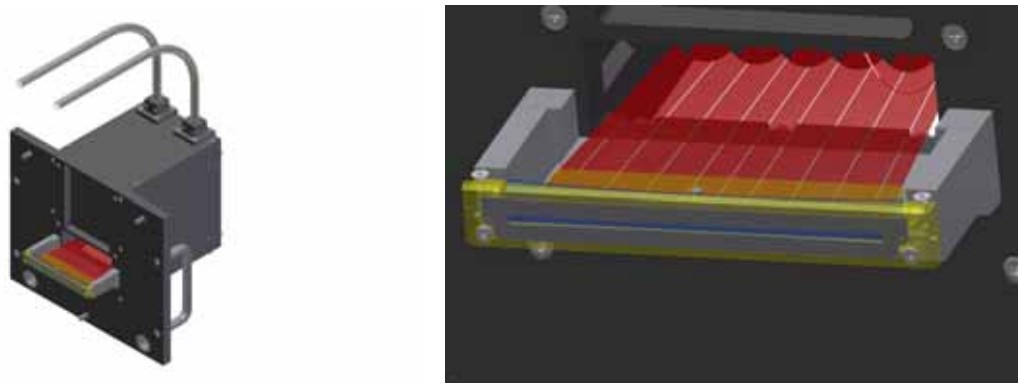

**Figure 5.** Left: CAD view of the slit-unit, which contains two exit ports for cables (for science and calibration fibers) and a precision reference plate to be attached to the collimator. Right: Zoomed view of the fiber-slit and lens cell. Note the narrow aperture mask in front of the lens to baffle stray light.

## 4. MANUFACTURE AND ASSEMBLY OF THE PROTOTYPE

Most mechanical parts, including all the critical components, for the VIRUS-IFU were manufactured at the mechanical workshop of the Astrophysical Institute Potsdam (AIP). Some standard items were outsourced to local companies. The fibers themselves were bought from two suppliers and the assembly was done at the AIP laboratories.

### 4.1 Test-assemblies

Various mini-IFUs (with 4x4 fibers) were assembled prior the manufacture of the actual prototype. Two different approaches were investigated. Firstly, a precision aperture mask was drilled in the mechanical workshop and filled with fibers (Fig. 6, left). While the position accuracy of the mask met the requirements, the alignment of the optical axes of the fibers (being perpendicular to the mask surface) was problematic to achieve. Also, the thinness of the mask, the sharpness of the holes (when using a steel mask) that damaged the fiber buffers or smearing effects during polishing of the surface (when using a plastic mask) were of concern. Alternatively, a U-shaped brass mount was filled with empty capillary tubing into which the fibers were inserted individually (Fig. 6, right). As the capillaries are 20mm long and have an inner diameter to match the fiber outer diameter, the parallel alignment of the optical axes was much better. Also, the robustness and the polishing of the glued array were satisfactory. However, the lateral position accuracy depended on the overall tolerances of the mount and needed to be carefully designed. For the manufacture of the prototype the capillary tube approach was chosen, while the assembly using a position mask was realized for a commercially produced bundle (see §5).

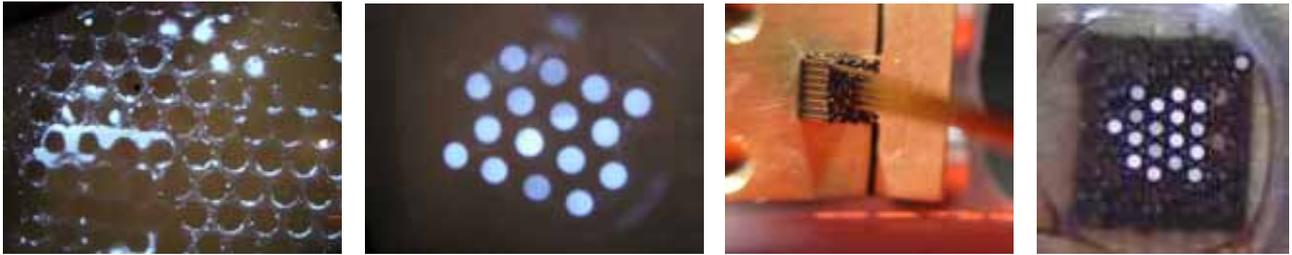

**Figure 6.** Photos of test assemblies, using either a drilled aperture mask (left) or a mount filled with empty capillaries (right) to insert the optical fibers.

### 4.2 The fiber cable

Contrary to PMAS[11,10], where the fibers are housed in several sub-cables[13] for protective purposes, both size and weight restrictions forbid this for VIRUS. While there is enough space at the slit-end to accommodate many cables, at the input end, the IFU head, its mount and subsequently the following cable needs to observe the size restrictions of 9 mm as to allow the tight packing fraction. Also, as there may be as many as 150 fiber bundles with a length of 15 meters in the full VIRUS configuration, weight load on the HET tracker becomes an issue. Therefore, all 246 fibers of one bundle will be housed and routed within one cable. The cable consists of an inner plastic fabric to ensure a frictionless movement of the fibers. This is surrounded by tubing made of a steel spiral with a plastic coating, which provided the protection against pressure, pull and strong bending.

For the prototype, the length of the fibers and was decided to be 4.5 meters. The protective cable is 4 meters long, as half a meter of bare fibers is located within the slit-unit. Fibers of the type FBP200220240 (core/cladding/buffer sizes: 200/220/240 microns) from Polymicro Technologies LLC (www.polymicro.com) were selected for the prototype, mainly because of their broadband transmission performance and previous experience with fibers from this manufacturer. For details with respect to FBP transmission properties see Haynes at al.[14]. Fibers of this type and the AS200220UVPI fiber from FiberTech (www.fibertech.de) were ordered and tested for FRD properties in the laboratory at AIP (see §6).

### 4.3 The IFU head

The head mount essentially consists of two parts, a U-shaped bottom and an upper lid. The mount measures 9 x 9 mm on a side times 25 mm in length. In a first step, the bottom part of the mount was filled up, row-by-row, with capillaries, which extended the length of the mount on both sides (see Fig. 7, top left). Once the lateral alignment of the matrix was satisfactory, the lid was closed to secure the assembly (Fig. 7, top right). Then, the fibers were inserted into the empty capillaries, such that they extend beyond mount and capillaries (Fig. 7, bottom left). Afterwards, the entire assembly was dipped into EPO-TEK 301-2, a non-shrinking two-component epoxy from Polytek Inc. The capillary force pulled the epoxy throughout the mount and filled all inter-fiber gaps.

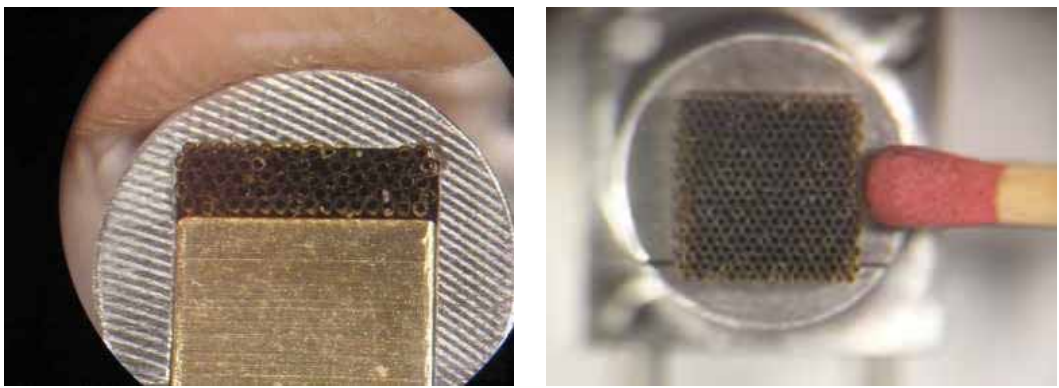

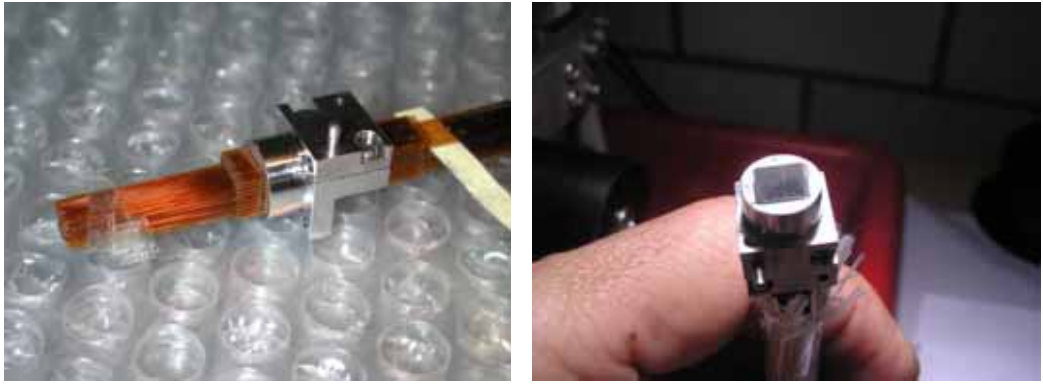

**Figure 7.** During the assembly of the IFU head, a precision mount was filled with empty ferules that act as spacers, and provide the required pitch. The assembly was filled with fibers, glued and polished. Note the size of the match to indicate the dimension.

After the glue set, the excess length of the fibers was cut and polished (Fig. 7, bottom right). Polishing was done using a laboratory set-up[15] with polishing sheets of grain sizes ranging from 63 to 3 microns. Thereafter, the plug was connected to a commercial polishing machine (from Buehler Inc.) and polished further, using sheets with grains sizes of 1 and 0.3 microns, until a smooth, flat and perpendicular surface was achieved. The progress was inspected repetitively, using a video microscope. Finally, an anti-reflective coated glass-plate (size: 6mm x 6mm x 1mm) was index-matched on top of the fiber-bundle.

### 4.4 The fiber-slit

In order to achieve the required position and pitch tolerances for the placement of the fibers, a monolithic block of stainless steel was cut with 255 v-grooves (Fig. 8). The layout is symmetrical about the center line. The v-grooves have a pitch of 320 microns and an angular pitch of 0.043 degrees from groove to groove. The total extent of the slit assembly from center to edge, along the circumference of the 420 mm spherical surface is 40.74 mm to the outer edge of the outer fiber. For reference, 200 microns is re-imaged to about five 15 micron pixels at the CCD. Inspection revealed that this critical piece could be manufactured to machine tolerances of a few microns, i.e. to less than 1% accuracy. In order to place and hold all fibers into their respective grooves, it was necessary to design and build an assembly tool as well. This assembly jig also ensured that the fiber ends terminate along a circumference of 420 mm. For the gluing of the fibers to the steel-block, the epoxy EPO-TEK 301-2 was used again. As there are more grooves (255) than science fibers (246), nine additional fibers were distributed across the entire slit, and can be connected to calibration light sources. This follows the same scheme as was successfully implemented for the PPak-IFU[10]. Apart from simultaneous calibration within a science exposure, these fibers also allow a careful investigation of any stray light features within the spectrograph or the analysis of the spectral profile at the detector (see 10, 11 for details).

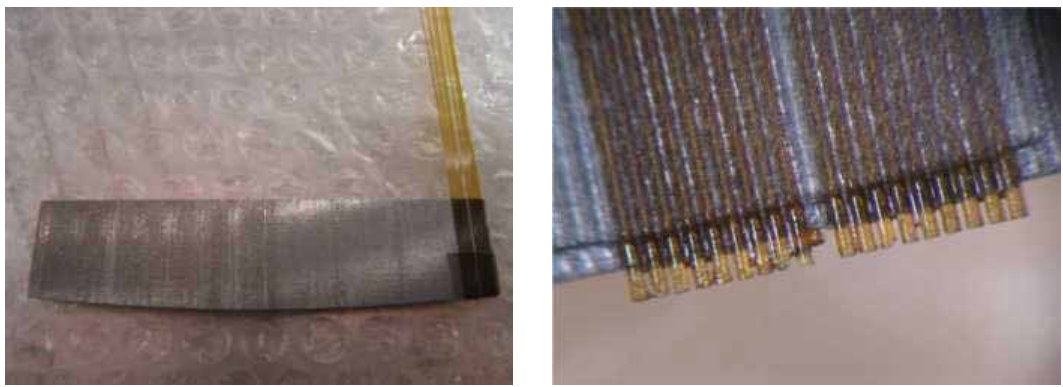

**Figure 8.** Photos of the monolithic slit substrate, with grooves for the fibers. The pitch from fiber center to center in the grooves is 320 microns, the overall slit length is 81.5 mm. For test purposes, a few (not perfectly polished) fibers were assembled.

## 5. COMMERCIAL BUNDLE

One of the major aims of the VIRUS development is to find design solutions to create relative simple subsystems that can be replicated and produced in a small-series by a commercial manufacturer, as to reduce the overall costs and manpower usually required for classical astronomical instrumentation. As part of a research and development (R&D) undertaking, a second fiber bundle was produced by the company FiberTech (www.FiberTech.de), Berlin. This bundle (Fig. 9) features 246 AS200220UVPI type fibers, with the required VIRUS input configuration. The approach by FiberTech to assemble the bundle was to use precision aperture masks to ensure accurate fiber positioning. A joint development between AIP and FiberTech is under way, to also find a commercially viable solution to manufacture the complex fiber slit. Furthermore, FiberTech provided fibers with custom-adjusted numerical aperture (NA) to match the f-number of the spectrograph. Both, the quality of the commercial built prototype bundle and the option to obtain custom-drawn fibers from this manufacturer are promising prospects for the full VIRUS manufacture, but need to be investigated further.

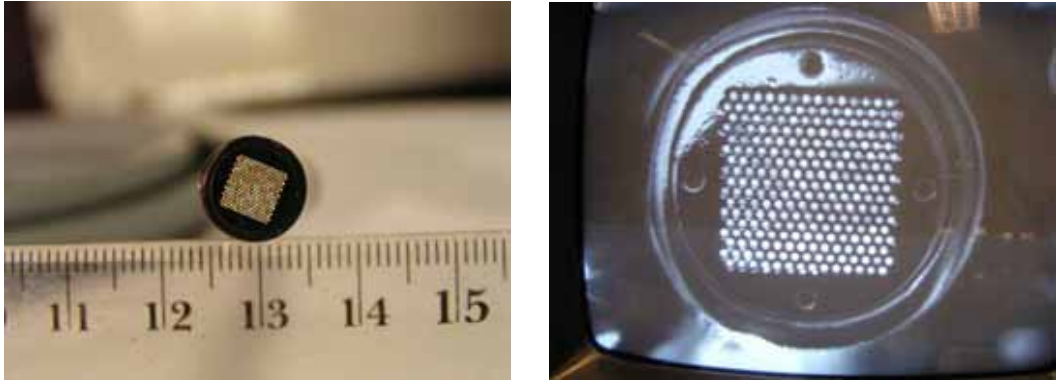

**Figure 9.** Photos of the second prototype, a fiber bundle that was built commercially by FiberTech, Germany.

## 6. FIBER TESTS

### 6.1 Fiber test-bench setup

Different kinds of individual fibers and assembled bundles were tested in the AIP laboratory, using a fiber test-bench that was originally built by Schmoll[16]. With this test-bench[17] it is possible to illuminate the fiber(s) with input beams of various f-numbers and with different light sources. The output light cone is projected onto a CCD detector, so that the profiles, the intensity, the output f-number, and thus FRD can be measured. The optical set-up is illustrated in Fig. 10.

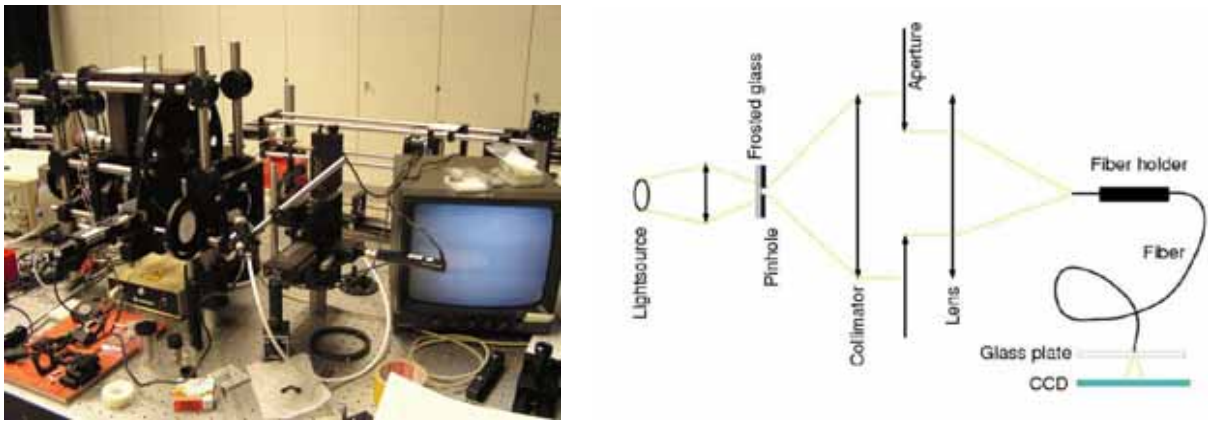

**Figure 10.** Left: The fiber test bench at AIP to measure fiber properties and FRD. Right: Illustration of the optical set-up.

## 6.2 Test Results

As the VIRUS bundle was still under development at the time of this paper, only a selection of preliminary results of the fiber tests are given here. For additional test results see Grupp[17].

Typically, quartz/quartz step-index fibers have a numerical aperture of NA=0.22. This corresponds to an f-number of approximately f/2.3. As this is faster than the beam that is used to feed the fibers, it is interesting to investigate, if a fiber NA that is tuned to match the output f-number is of advantage. In case of VIRUS the input beam is f/3.65 and the collimator can accept up to f/3.36. However, any rays faster than this will not be collimated, overfill subsequent optics and eventually be scattered and create stray light. A properly tuned NA of the fiber may therefore be used as a baffle, to avoid any light that is outside the collimator at the first place. The corresponding numerical aperture for a f/3.36 collimator optics is 0.15. Adding some tolerance, a AS200220UVPI fiber with customized NA=0.16 was produced at FiberTech. Fig. 11 shows the output light distribution of a fiber with NA=0.16 (left) and NA=0.22 (right), both were illuminated with a f/3.5 input beam. Note the fainter halo around the nominal light cone in the latter case. This halo is in between f/3.5 and f/2.3 and exactly what one wants to avoid, as done with the tuned NA-fiber.

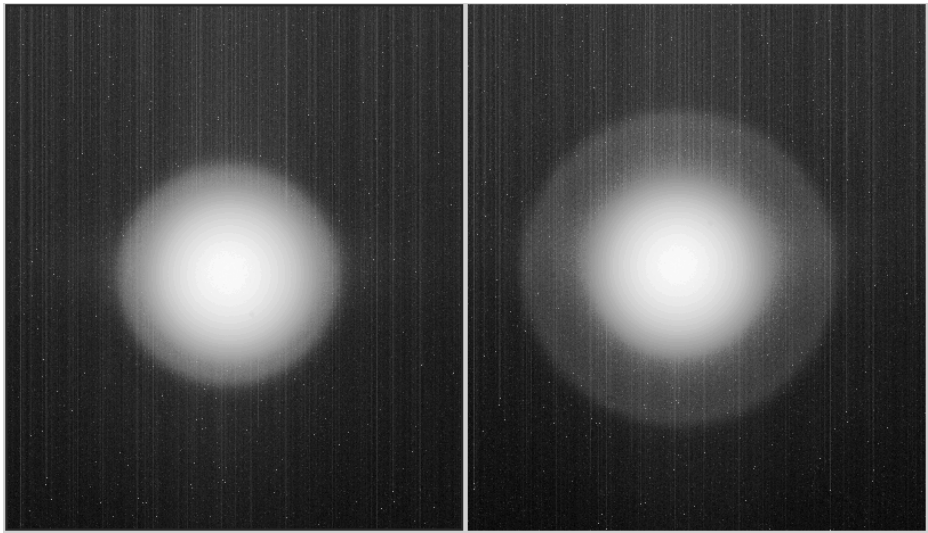

**Figure 11.** Comparison of the output light cone of a 'standard' fiber with NA=0.22 (right) to a fiber with 'tuned' NA of 0.16 (left). Given an input beam of f/3.5 (NA=0.14), the standard fiber shows an extended halo that is outside the nominal beam and may cause scattered light within the spectrograph. The tuned NA fiber suppresses this unwanted light.

As the placement of the spectrographs and therefore the length of the fiber bundles for the final VIRUS instrument are not fixed yet, efficiency and FRD properties were measured for a set of short (4.5 m) and long (14.5 m) fibers. The results in Fig. 12 (left) indicate, that FRD behavior is affected by length, but mainly for slower input f-numbers. Given the input coupling for VIRUS using f/3.65, FRD for long and short fibers seems to be identical. Obviously, this is not the case with respect to transmission, where shorter fibers are beneficial (Fig. 12, right. Also, refer to data sheets from manufacturers and measurements by Haynes et al.[14]).

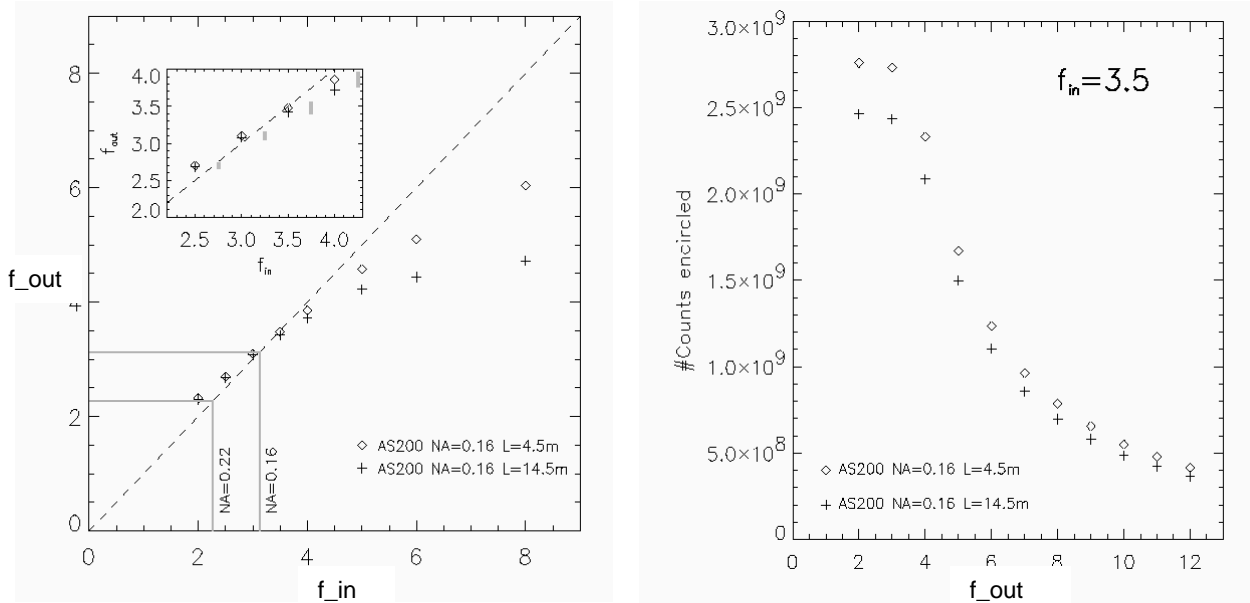

**Figure 12.** Left: Plot of output versus input f-number to compare FRD properties between a long (14 m) and a short (4 m) fiber. Right: Plot of encircled counts versus output f-number for a given input beam of f/3.5.

The fiber bundle for VIRUS is index-matched on both sides to either a anti-reflective coated glass plate or lens. Immersion not only reduces Fressnel losses, but is also beneficial with respect to FRD properties[13,16]. This finding was confirmed during measurements for the VIRUS prototype and is plotted in Fig. 13, left.

As the input beam given by HET features a central obstruction (and in fact is taken advantage of by the spectrograph design featuring a Schmidt system), this was simulated during the tests too. The resulting output of a f/3.5 beam with a central obstruction corresponding to f/8 is shown as double-peaked profile in Fig. 13, right. This profile is in agreement with a measured f/8 beam of which a f/3.5 beam was subtracted (see over-plotted profiles in Fig. 13, right).

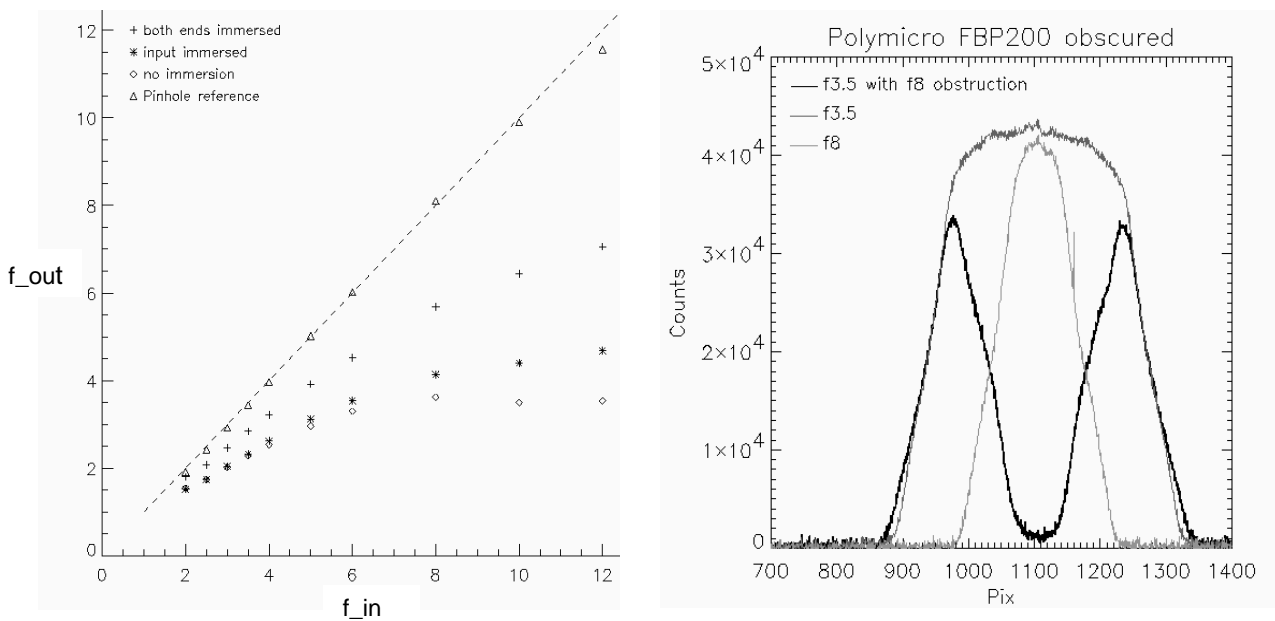

**Figure 13.** Left: plot of output versus input f-number to illustrate the beneficial effect of index-matching to minimize FRD effects. Right: the output for an input beam featuring a central obstruction is similar to folding beams with the appropriate f-numbers.

# 7. ONGOING WORK

At the time of writing, the fiber-bundle was not fully completed and further tests in the laboratory were ongoing. After commissioning of the VIRUS prototype, foreseen for August 2006 at the 2.7m Harlan Smith Telescope at McDonald observatory, an evaluation of the on-sky performance and the critical interplay[18] between fibers and spectrograph projection will be possible. If the requirements are met, the strategy for the full VIRUS development, featuring some 145 spectrographs and integral-field units, implies technology transfer to industrial partners to allow a commercial replication of these subsystems at reasonable costs.

# ACKNOWLEDGEMENTS

We thank the workshop at AIP for the careful manufacture of the mechanical parts of the IFU. The use of fused silica capillary tubes to establish the layout of fibers was suggested by Gary Nelson at Polymicro Technologies Inc. We thank Mr Heyer and Mr Zoheidi for valuable advises that resulted in the production of the fiber-bundle at FiberTech. Helpful discussions with Jeremy Allington-Smith, Jürgen Schmoll, and Matt Bershady are gratefully acknowledged. The VIRUS prototype is partially funded by a gift from the George and Cynthia Mitchell Foundation, by AFRL under agreement number FA9451-04-2-0355, by the Texas Advanced Research Program under grant 003658-0005-2006, the McDonald Observatory, and by the Max Planck Institut fuer Extraterrestriche-Physik (MPE). The IFU for the VIRUS prototype is a contribution by the Astrophysikalisches Institut Potsdam (AIP).